\def\be{\begin{equation}}
\def\ee{\end{equation}}
\documentstyle[preprint,aps,bezier,eqsecnum]{revtex}
\begin{document}
\draft

\preprint{FERMILAB-Pub-93/364-A}

\title{Gravitational Lens Time Delays and
Gravitational Waves}
\author{Joshua A. Frieman$^{1,2}$, Diego D. Harari$^3$, and
Gabriela C. Surpi$^3$}
\address{${}^1$NASA/Fermilab Astrophysics Center \\
Fermi National Accelerator Laboratory \\
P.O. Box 500, Batavia, IL 60510, USA}
\address{
${}^2$Department of Astronomy and Astrophysics \\
University of Chicago, Chicago, IL 60637}
\address{${}^3$Departamento de F{\'\i}sica,
Facultad de Ciencias Exactas y Naturales\\
Universidad de Buenos Aires \\
Ciudad Universitaria - Pab. 1,
1428 Buenos Aires, Argentina}
\maketitle
\begin{abstract}
\tighten
Using Fermat's principle, we analyze the effects of very long
wavelength gravitational waves upon the images of a gravitationally
lensed quasar. We show that the lens equation in the presence of
gravity waves is equivalent to that of a lens with different alignment
between source, deflector, and observer in the absence of gravity
waves. Contrary to a recent claim, we
conclude that measurements of time delays in gravitational lenses
cannot serve as a method to detect or constrain a stochastic
background of gravitational
waves of cosmological wavelengths, because the wave-induced time
delay is observationally
indistinguishable from an intrinsic time delay due to
the lens geometry.
\end{abstract}
\pacs{PACS numbers:  98.80.-k, 04.80.+z, 95.30.Sf}

\section{INTRODUCTION}
A stochastic background of gravitational waves of
cosmological wavelengths may arise in the early Universe, for
instance as a consequence of
quantum effects during a period of inflationary expansion, or as the
result of gravitational radiation by oscillating
cosmic strings. Its presence could be manifested as a large angular
scale anisotropy in the cosmic
microwave background, induced by the Sachs-Wolfe
effect\cite{Sachs67}, the differential
redshifting of photons in the presence of tensor metric perturbations.
It is possible that a significant
fraction of the anisotropy measured by the COBE DMR
experiment\cite{COBE} is due to cosmological gravitational
waves.\cite{COBEGW}  So far, the
microwave observations cannot determine how much of the
anisotropy is due to tensor perturbations (gravitational waves) and
how much to scalar (energy-density) fluctuations.

Another potential method to reveal the presence of gravitational
waves of cosmological
wavelengths was recently suggested by
Allen,\cite{Allen89,Allen90} namely, to use
measured time delays between gravitationally lensed multiple
images of distant quasars.
Gravitationally lensed multiple images of a source such as a quasar
arrive at the Earth at different times if the source, deflector (the
lensing body), and observer are not
in perfect alignment, because there is a
difference in geometric path lengths between, and in the deflector's
gravitational
potential traversed by, the different light rays. We shall call these
two effects the `intrinsic' time
delay of the lens. For a lens geometry where
$L$ is the distance between observer and deflector and $2\eta$ is
the angular separation between the images,
the typical intrinsic time delay is $\Delta T \sim
L \eta^2$. Any `extrinsic' perturbation to the spacetime metric
(i.e., not associated with the lens itself) would be
expected to cause additional time delays between the images. For
example, an additional time delay
would be induced by a cosmological background of very long wavelength
gravitational waves. \cite{Allen89,Allen90} As shown by Allen, for
waves with frequency $\omega \sim L^{-1}$, the gravity wave-induced
time delay is of order $\Delta T\approx L h\eta$, where $h$ is the
dimensionless amplitude of the gravitational wave. Therefore,
waves of amplitude $h \agt \eta$ would be expected to have drastic
effects on lens time delays.

Based on this effect, Allen claimed that gravitational
lenses could serve as gravitational wave detectors,
\cite{Allen89,Allen90} and that a
bound could be placed on the amplitude of
gravitational waves of cosmological wavelengths from the
requirement that the wave-induced time delay in the double quasar
0957+561 not exceed the observed delay of 1.48 yr. \cite{0957}
(The observed delay is generally attributed to the intrinsic
delay of the lens.)
For 0957+561, the image angular separation is $3\times 10^{-5}$
rad $= 6.1$ arcsec, and Allen obtained the bound $h<2\times 10^{-5}$.
A gravity-wave background which nearly saturates the above bound would
have important implications for gravitational lens models and would
seriously compromise attempts to
use lens time delays to measure the Hubble parameter.

Subsequent to Allen's work, the microwave anisotropy bound
on the amplitude of cosmological gravitational waves has been
significantly tightened by COBE. Through the Sachs-Wolfe
effect, gravitational waves induce a temperature anisotropy of order
their dimensionless amplitude. From the COBE detection of the
quadrupole anisotropy, it follows that $h\alt (\delta
T/T)_{\ell = 2}\approx 6\times 10^{-6}$ for wavelengths comparable to
the present Hubble radius,
$\lambda \sim H_0^{-1} = 3000$ h$^{-1}$ Mpc.
This bound is roughly a factor of three smaller than Allen's limit.
However, although 0957+561 is the first
gravitational lens system for which a time delay has been reliably
measured, other lens systems are
also being monitored; in particular, for a lens
with smaller image angular separation $\eta$ and thus smaller
intrinsic time delay, the wave-induced delay would be even more
important, and the
corresponding lens bound on $h$ potentially more restrictive.
Inflationary models suggest that a significant fraction
of the quadrupole anisotropy could be
due to gravitational waves \cite{COBEGW,Abbott84}. If this is the
case, then time delays induced by
gravitational waves in gravitationally lensed quasars would be
significant.

In this paper, we reconsider Allen's proposal. Our central theme is
that, for measurements of time delays in gravitational lenses to serve
as gravitational wave detectors,
the observer must be able to separate the
wave-induced time delay from the intrinsic time delay originating in the
lens geometry. We discuss the feasibility of
observationally distinguishing these two sources of time delay.
We approach this issue through application of Fermat's principle, a
useful tool for analysing gravitational lens problems
\cite{Blandford86,Schneider92}, which has recently been shown to
hold in the non-stationary space-times we
consider \cite{Kovner90,Nityananda92}. We conclude that
measurements of time delays in gravitational lenses are not likely to
serve as a method to detect or constrain a cosmological background
of gravitational waves, because the wave-induced time delay is
observationally indistinguishable from the intrinsic time delay of an
alternative lens geometry.
As a consequence, the cosmological applications of lens time delays,
e.g., inferring $H_0$ or galaxy masses, are not affected by gravity
waves, regardless of the amplitude $h$.
We note that, using quite different methods, the same
conclusions were reached for general (scalar, vector, and
tensor) metric perturbations
by Frieman, Kaiser, and Turner. \cite{FKT}

In Refs. \cite{Allen89,Allen90}, the time delay induced by a
gravitational wave upon a gravitational lens was evaluated through
the Sachs-Wolfe formula
\cite{Sachs67} for the differential photon
redshift in the presence of metric fluctuations, integrated along
unperturbed photon paths, {\it i.e.}, along the same trajectories the
photons would have followed in the absence of the wave. As we will
show, this method is not applicable in the case that the wave amplitude $h$
is comparable to or larger than the angular separation $2\eta$ that the
images would have in the absence of the wave, and the expression
for the time 
delay derived in Refs.
\cite{Allen89,Allen90} is valid only if $h \ll \eta$. In the opposite
limit, $h \gg \eta$, the effect of the wave is equivalent to a change in
the alignment of the system so large that multiple images do not
form (at least for non-singular lens potentials).
Thus, the wave-induced time delays never exceed typical intrinsic
delays, and cannot be used to constrain the amplitude of cosmological
gravitational waves. Moreover, even in cases where the wave-induced
delay is comparable to the typical intrinsic delay, we will show that
an observer would attribute the entire delay to the intrinsic lens
geometry. Thus, the wave-induced delay cannot be unearthed in
practice or in principle.

To address these issues, we explicitly take into account the spatial
distortion of the photon trajectories
induced by the gravitational waves, which
is non-negligible even if $h<<\eta$. The wave-induced
perturbation of the photon paths gives rise to extra
contributions to the time delay, in addition to the differential
redshift along the two trajectories. The extra contributions arise as
a consequence of difference in path lengths and different
gravitational potential traversed by each photon due to the
asymmetry in their trajectories induced by the wave. When the dust
settles, our result for the wave-induced time delay coincides
with that of Refs. \cite{Allen89,Allen90} because these extra terms
cancel each
other, but only in the limits $h\ll \eta$ and  $\omega L\eta \ll 1$.
Moreover, the spatial distortion of the photon paths is always
very significant when it comes to the interpretation of lens observations:
a gravitational wave distorts the apparent angular positions of the
images relative to the deflector in just such a way that an observer
would attribute the wave-induced time delay to
an intrinsic time delay associated with the image-deflector
misalignment he or she sees.
Since the lens geometry is not known {\it a
priori}, but reconstructed from observations, one could equally well
adjust the measurements to a given lens geometry in the presence of
gravitational waves, or to an alternative lens geometry and no waves
at all. Thus, it appears observationally impossible to distinguish
wave-induced time delays from intrinsic delays, and so to
detect cosmological gravitational waves through time delay
measurements in gravitationally lensed quasars.

\section{TIME DELAY IN A SIMPLE LENS CONFIGURATION}

To more clearly display the features discussed above, we first analyze
a simple lens model: a Schwarzschild (point mass) lens in a highly
symmetric configuration, and a gravitational wave propagating
perpendicular to the lens axis. In the next section we generalize the
results derived here to the case of an arbitrary thin lens and
arbitrary polarization and wave vector of the gravitational wave.
Consider a static, spherical body of mass M, located at the origin
of coordinates, that deflects photons emitted by a point-like source
located at $(x=0,y=0,z=-L)$, and an observer on the extension of
the source-deflector line at $(x=0,y=0,z=+L)$ (see Fig. 1).
Given the axial symmetry, the observer sees an Einstein ring image
of the source, with angular radius $\eta\equiv \sqrt{2GM/L}$.
Here, $G$ is Newton's
constant, we take the speed of light $c=1$, and we assume $\eta <<1$.
To simplify the discussion, we focus on those
photons that travel along the $y=0$ plane, forming two images on
opposite sides of the ring. In the absence of a gravitational wave,
there is no time delay between the
two images, and they arrive with an angular
separation $\Delta \theta = 2\eta$. Now consider a gravitational
wave of dimensionless amplitude $h$ and
frequency $\omega$, with polarization (+), propagating along the
$x$ axis in the positive $x$ direction. Sufficiently far from the
deflector mass, the spacetime interval can be approximated by
\begin{equation}
ds^2=\left(1-{2GM\over r}\right)dt^2-\left(1+{2GM\over r}\right)
(dx^2+dy^2+dz^2)+ h\cos\omega(t-x)(dy^2-dz^2)\  .\label{ds1}
\end{equation}
Along a photon path, $ds^2=0$. Thus, if the spatial photon
trajectories $x=x(z)$ were known, one could evaluate the time of
travel by simple integration in $z$ from $-L$ to $L$,
\begin{equation}
T\approx\int_{-L}^Ldz \left[1+{1\over 2} \left({dx\over
dz}\right)^2+{1\over 2}h\cos\omega(t-x) +{2GM\over r}\right]\  .
\label{T}
\end{equation}
To the level of approximation we shall be working (we are interested in
terms of order $h\eta$ in the time delay), $t$ can be replaced
in eq.\ (\ref{T}) by $t=t_e+(z+L)$, with $t_e$ the time at which the
photons were emitted at $(x=0,z=-L)$. The first two terms in the
integrand of eq.\ (\ref{T}) are the geometric
contribution to the time of travel, while the third and fourth are
contributions from the gravitational potential of the wave and the
deflector respectively.

\subsection{Integration along unperturbed paths}

Let us first evaluate the time delay along unperturbed photon
trajectories. We approximate each path by straight
segments, deflected by angle $\alpha=2\eta$ at the deflector plane
$z=0$ (see Fig. 1). The
approximation by straight segments is convenient and
appropriate in the case of thin gravitational lenses, where most of
the deflection occurs in the immediate vicinity of the deflector
plane \cite{Schneider92}. The light trajectories are then
\begin{eqnarray}
x_{1,2}=&&\pm\eta(z+L)\quad\quad z<0\nonumber\\
x_{1,2}=&&\mp\eta(z-L)\quad\quad z>0\label{xzunp}
\end{eqnarray}
where the subscript (1,2) distinguishes trajectories that pass along
opposite sides of the deflector.
Straightforward integration leads to the time of travel, $T_{1,2}$.
Clearly, given the symmetry of the integration paths, the only
contribution to the time delay comes from the different gravitational
wave potential encountered by each trajectory, {\it i.e.}, from the
third term in the integrand of eq.\ (\ref{T}). The time delay is
\begin{equation}
\Delta T=T_1-T_2=-{h\eta\over\omega}[\sin\omega(t_e+2L)+
\sin\omega t_e-2\sin\omega(t_e+L)\cos\omega L\eta]\ ,\label{dt1}
\end{equation}
which, if $\omega L\eta <<1$, is approximated by
\begin{equation}
\Delta T\approx 4{h\eta\over\omega}\sin^2 \left({\omega L\over 2}
\right) \sin\omega(t_e+L)\
.\label{a}
\end{equation}
This coincides with the result of Ref. \cite{Allen89}, evaluated by
integration of the Sachs-Wolfe formula along the same unperturbed
trajectories.

The method used above to evaluate the time delay induced by the
gravitational wave is questionable, even if $h<<\eta$, because the
actual photon trajectories are perturbed by the wave, and are
expected to be neither straight nor
symmetric with respect to the lens axis. As a result,
one path may have smaller impact parameter with respect to the
deflector than the other, and
hence be deflected by a larger angle. This asymmetry in the paths leads
to differences in both the geometric and potential
contributions to the time of travel as large as that evaluated above.
We shall evaluate these extra contributions, after derivation of the lens
equation through Fermat's principle, and find that they
cancel to leading order only if $h \ll \eta$. Thus, eq.\ (\ref{a}) is only
valid in this limit (moreover, in this limit, the wave-induced delay
would be swamped by the intrinsic delay if the perfectly aligned
symmetric lens were replaced by a
misaligned lens such as that in Fig. 2). Furthermore,
we will show that, even in this limit, the delay (\ref{a}) would be
attributed by the observer to intrinsic lens delay, that is, to
an apparent misalignment between source, deflector, and observer.

To prepare for this result,
we first briefly review the derivation of the lens
equation and time delay for a misaligned lens in the absence of
gravity waves.

\subsection{Fermat's principle for a Schwarzschild lens}

Fermat's principle provides a useful shortcut in many lensing
problems: one can approximate
the photon trajectories by null zig-zag trial paths and then extremize
the time of travel \cite{Blandford86,Schneider92}, instead of solving
the geodesic equations. Consider the lens
depicted in Fig. 2. The source is at an angle
$\beta$ with respect to the line that joins observer and deflector,
and in this subsection we assume there is no gravitational wave present.
Consider a path that, starting from the source at $z=-L$, moves along
a straight line up to the $z=0$
plane, where it is deflected by an angle $\alpha$,
and then arrives at the observer at $z=L$, forming an angle $\theta$
with respect to the line that joins observer and deflector. The angles
$\alpha, \beta$ and $\theta$ must satisfy
\begin{equation}
\alpha=2(\theta -\beta)\  . \label{alfa}
\end{equation}
Assuming $\beta,\theta<<1$,
the time of travel along such a null path would be
\begin{equation} T\approx 2L+\theta^2L-2\beta \theta
L-4GM\ln\theta\  , \label{t1}
\end{equation}
where we have neglected constant ($\theta$-independent) terms.
The first three terms are of
purely geometric origin, while the last originates in the
gravitational potential of the deflector. The condition that $T$ be
an extremum ($dT/d\theta=0$ for fixed $\beta$) gives
\begin{equation}
\theta^2-\beta\theta-{2GM\over L}=0\ , \label{le}
\end{equation}
which is usually referred to as the lens equation \cite{Schneider92}.
The solutions give the angular positions of the two images:
\begin{equation}
\theta_{1,2}={1\over 2}(\beta\pm\sqrt{\beta^2+4\eta^2})
\quad\quad {\rm with}\quad\quad \eta^2\equiv{2GM\over L}\
.\label{12}
\end{equation}
Different signs indicate that the images appear on opposite sides of
the deflector. The resulting time delay is given by
\begin{equation}
\Delta T=T_1-T_2=(\theta_1^2-\theta_2^2)L-2\beta (\theta_1-
\theta_2)L-4GM\ln(\theta_1/|\theta_2|)\ . \label{td1}
\end{equation}
In the limit of small misalignment angle, $\beta \ll \eta$, the
image angular positions are approximately
\begin{equation}
\theta_{1,2}\approx
\pm\eta+{1\over 2}\beta
\ .\label{12ap}
\end{equation}
and the time delay reduces to
\begin{equation}
\Delta T \approx -4\beta\eta L \approx -2\beta \Delta \theta \ ,
\label{td1a}
\end{equation}
where $\Delta \theta = \theta_1 - \theta_2$ is the image angular
separation.

For the arguments we will make below,
it is useful to bear in mind how lens observations in such a simple
system could be used to extract cosmological information.
If the deflector is seen in addition to the two images, then the lens
observables are $\theta_1$, $\theta_2$, and $\Delta T$. The observer can
then infer the misalignment angle from $\beta = \theta_1 + \theta_2$,
and the lens parameter $\eta$ from eq.\ (\ref{12}). Using these
observed and derived quantities, eq.\ (\ref{td1})
can be used to determine $L$. (More generally, if
the source and observer are not equidistant from the lens, the
reasoning above determines a
distance measure for the lens.) Comparison with the
deflector redshift then yields an estimate of the Hubble parameter
$H_0$. For deflectors more
complex than point masses, the observables
above must be supplemented by information about the deflector
potential obtained, e.g., from
measurement of the lens' velocity dispersion.

\subsection{A Schwarzschild lens with a gravitational wave}

Now we proceed with a similar technique, based on Fermat's principle,
to evaluate the time delay induced by a gravitational wave in the
symmetric Schwarzschild lens configuration, with source, deflector,
and observer aligned as in Fig. 1.
The validity of Fermat's principle in
non-stationary spacetimes was recently discussed in the context of
gravitational lensing problems \cite{Kovner90,Nityananda92}.
We evaluate the time of travel, integrating eq.\ (\ref{T}).
Instead of integrating along straight lines, however, we take the two
segments of each zig-zag trial path to be
null geodesics of the gravitational wave metric,
as they would have been in the absence of the lensing body.
We work up to the order of
approximation needed to study terms of order $h\eta$ in the time delay,
and we also assume $\omega L\eta<<1$. This is the most interesting
range, since the effect under consideration becomes largest
when $\omega L\approx 1$.
The geodesic equations in the metric\ (\ref{ds1}), with $M=0$, lead
to
\begin{equation}
{dx\over dz}\approx\gamma -{1\over 2}h \cos\omega(t_e+z+L)
\  .\label{dxdz1}
\end{equation}
Here, $\gamma$ is an arbitrary integration constant, the average
slope of the trajectory, assumed to be small.
Note that, as before, in the
argument of the cosine in\ (\ref{dxdz1}), we have replaced $t$ by
$t_e+z+L$; we have also dropped the dependence on $x$
because it is unnecessary to include it
to evaluate the time delay to order $h\eta$, in the limit $\omega
L\eta <<1$. The third term in eq.\
(\ref{T}) is the only one where the $x$-dependence
inside the cosine needs to be included, and it is enough to do so at
zero order.

Now we choose the integration constants so that a
trajectory that starts from $x=0,z=-L$ at $t=t_e$, and deflected
by an arbitrary angle in the $z=0$ plane, arrives at the observer at
$x=0,z=L$. One finds that
\begin{eqnarray}
{dx\over dz}=&&\ \epsilon - {1\over 2}h\cos\omega(t_e+z+L)\quad
{\rm if}\quad z<0\nonumber\\
{dx\over dz}=&&-\epsilon +{h\over 2\omega L}[\sin\omega(t_e+2L)-
\sin\omega t_e]- {1\over 2}h\cos\omega(t_e+z+L)\quad{\rm
if}\quad z>0
\ .\label{xzp}
\end{eqnarray}
Here, $\epsilon$ is an integration constant which parametrizes
the family of trajectories that meet
the focusing conditions at the required points. Each trajectory
consists of two segments which are null geodesics of the gravitational
wave metric, neglecting the deflector potential, which will be taken
into account through Fermat's principle.
According to the latter, the actual trajectories
are those null paths that extremize the time of travel with respect to
variations of the parameter $\epsilon$ in the metric that
includes both the gravitational wave as well as the deflector's
potential.

Before we proceed to extremize, however, we parametrize the
trajectories in a different way, defining a parameter more
relevant to observations.
Instead of $\epsilon$, we use the angular position
of the image, which we denote by $\theta$, relative to
the angular position of the deflector, at the time
of arrival of the images at the observer.
We again use equation\ (\ref{dxdz1}) and fix the appropriate
integration constants to
determine the slope of the trajectory
of a photon that arrives from the deflector (i.e., the angular position
of the deflector),
which we denote by $dx_{\rm lens}/dz$,
\begin{equation}
{dx_{\rm lens}\over dz}\bigg|_{z=L}={h\over 2\omega
L}[\sin\omega(t_e+2L)
-\sin\omega(t_e+L)]-{1\over 2}h\cos\omega(t_e+2L)\ .\label{xl}
\end{equation}
Then the angular position $\theta$ of the image with respect to the
apparent deflector position is
\begin{equation}
\theta=-{dx\over dz}\bigg|_{z=L}+{dx_{\rm lens}\over
dz}\bigg|_{z=L}= \epsilon +
{h\over 2\omega L}[\sin\omega t_e-\sin\omega(t_e+L)]\  .
\label{theta}
\end{equation}
This relates $\theta$ to the parameter $\epsilon$ of eq. (\ref{xzp}).
Now the deflection imprinted by the lens upon the trajectory at $z=0$,
which we denote by $\alpha$, can be written as
\begin{equation}
\alpha\equiv {dx\over dz}\bigg|_{z=0^-}-{dx\over dz}\bigg|_{z=0^+}=
2(\theta-\beta_g)\label{def1}
\end{equation}
where we have defined
\begin{equation}
\beta_g\equiv{h\over 4\omega L}[\sin\omega(t_e+2L)
+\sin\omega t_e-2\sin\omega(t_e+L)]=-
{h\over\omega L}\sin^2 \left({\omega L\over 2}\right)
\sin\omega(t_e+L)\  .\label{beta}
\end{equation}
We have defined the quantity $\beta_g$ in such a way that
the expression (\ref{def1}) for the deflection has the same form
as eq. (\ref{alfa})--in that case, $\beta$ measured the
misalignment between deflector,
source, and observer in the absence of a gravitational wave,
as in Fig. 2. We will see in what follows that in all
respects $\beta_g$ plays exactly the same effective role here.

Next we evaluate the time of travel, integrating eq. (\ref{T})
along the null trajectories (\ref{xzp}),
parametrized in terms of $\theta$, and find
\begin{equation}
T\approx 2L+\theta^2L-2\beta_g\theta L-4GM\ln\theta\  . \label{t2}
\end{equation}
As advertised, this has exactly the form of eq. (\ref{t1}), which
gave the time of travel for a similar lens with no gravitational wave,
but with lens and source misaligned by an angle $\beta$, as in Fig. 2.
Recall that in that case, the first three terms were of geometric
origin. In the present case, only the first two terms are geometric
(they come from  integration of $[1+(dx/dz)^2/2]$ in (\ref{T})).
The third term, proportional to $\beta_g$, is due to
the wave gravitational potential, and comes from integration of the
third term in eq.\ (\ref{T}). Finally,
the last term is due to the deflector's gravitational potential.

The equivalence of expressions  (\ref{t1}) and\ (\ref{t2})
leads to our main conclusion: the lens equation in the presence of
a gravitational wave is, to the order of approximation considered,
completely equivalent to that of a similar lens with a different
alignment and no gravity wave. The effective misalignment angle
$\beta_g$ is given by eq.\
(\ref{beta}) in terms of the wave parameters. The analogy is exact
only in the limit $\omega L\eta \ll
1$, but this is the interesting range in any
case, because $\eta \ll 1$ and time delays are largest for $\omega
L\approx 1$. Since
$\beta_g$ depends upon time, the analogy is only valid over periods
of time much shorter than
$\omega^{-1}$. Again, since the effect is
relevant only for waves of cosmological wavelengths, this
time-variation of the time delay is observationally irrelevant.

{}From eq.\ (\ref{t2}),
the time delay between the two images is given by expression (\ref{td1})
with the substitution $\beta \rightarrow \beta_g$,
\begin{equation}
\Delta T=T_1-T_2=(\theta_1^2-\theta_2^2)L-2\beta_g (\theta_1-
\theta_2)L
-4GM\ln(\theta_1/|\theta_2|)  \ . \label{tdx}
\end{equation}
So far, we have made no assumption about the relative amplitudes of
the gravitational wave effect,
$\beta_g \sim h$, and the deflector Einstein ring angular radius
$\eta$ (Cf. eq.\ (\ref{12}) ). However, if $\beta_g \gg \eta$,
the effect of  the wave is equivalent to
that of a system very much out of alignment. In this limit, the
magnification of the second image goes to zero as $(\eta/\beta_g)^4$, and
multiple image formation effectively does not take place.

In the opposite limit, $\beta_g \ll \eta$,  Fermat's
principle leads to the same result as
eq.\ (\ref{td1a}), but with $\beta$ replaced by the effective
$\beta_g$ of eq.\ (\ref{beta}),
\begin{equation}
\Delta T
\approx -4\beta_g\eta L=4{h\eta\over\omega}\sin^2 \left({\omega
L\over 2}\right)
\sin\omega(t_e+L)\
.\label{td2}
\end{equation}
This result coincides with that of eq.\ (\ref{a}), obtained through
integration
along unperturbed paths,
and is just Allen's result \cite{Allen89}.
Note that the additional term originating in a path length
difference,  $(\theta_1^2-\theta_2^2)L$, cancels the term
due to the deflector's gravitational potential, $4GM\ln
(\theta_1/|\theta_2|)$. The wave-induced distortion of the photon
paths is not negligible, however, when it comes to interpreting the
result.

Indeed, suppose the observer of this lens has no knowledge of the
possible existence of gravitational waves, and seeks to measure, e.g.,
the deflector mass $M$ or the Hubble parameter $H_0$ from her
observations.
The effect of the gravity wave upon the apparent angular
positions of the images and the deflector
would trick the observer into believing he or
she sees an ordinary misaligned lens.
Moreover, the observer's inference of the misalignment angle $\beta_g$ from
the observed image angular positions, $\beta_g = \theta_1 + \theta_2$,
and the observed image time delay would all be
in accord with this belief, and he or
she will infer the {\it correct} values for
$M$ and $H_0$, even though taking no account of gravitational waves
and instead assuming a homogeneous and isotropic spacetime (aside from
the deflector). That is, while the gravity wave does cause a time delay, it
covers its own tracks in a misalignment change, leaving no
measurable trace of its presence, and can be safely and consistently
ignored by the
lens observer. Thus it appears impossible to use time delay measurements
to detect cosmological gravitational waves even in principle.

For completeness, we emphasize that this conclusion holds
even if $\beta \agt \eta$, but that it is only in the
limit $\beta_g \ll \eta$ that the time delay agrees with eq.\ (\ref{a}).

\section{TIME DELAY IN A THIN LENS WITH GRAVITATIONAL WAVES}

In this section we show how the conclusions reached above can be
generalized to the case of an
arbitrary thin gravitational lens with gravitational
waves of arbitrary polarization and direction of propagation.
First we briefly review the features of a general lens when no
gravitational waves are present.
We assume, as usually applies for cases of astrophysical
interest, a thin, stationary gravitational lens, such
that the weak field approximation is valid \cite{Schneider92}.
Consider a lens geometry as in Fig. 2, only now we do not
assume that the photon paths
lie in the plane that contains source, deflector,
and observer:  $\vec\theta $ and $\vec\beta$  are two-component
angular vectors, that give angular positions at the observer's
location, $\vec\alpha $  is the deflection, and $\vec\xi=L\vec\theta$
determines the impact parameter in the deflector plane.
The condition that the photons from the source reach the
observer implies the following relation, the vectorial generalization
of eq.\ (\ref{alfa}):
\begin{equation}
\vec\alpha =2\ (\vec\theta -\vec\beta)\  , \label{G}
\end{equation}
and the time of travel is given by:
\begin{equation}
T\approx 2L +|\vec\theta|^2\ L-2\ \vec\beta\cdot \vec\theta \ L-
\psi(\vec\xi )\  ,\label{T1}
\end{equation}
which is the generalization of eq. (\ref{t1}). The last term
originates in the deflector gravitational potential, and is
given, for a thin lens, by
\begin{equation}
\psi (\vec\xi )=4G\int d^2\xi'\ \Sigma (\vec\xi ')
\ln \left({\vert \vec\xi -\vec\xi '\vert\over\xi _0}\right)\  ,
\label{PSI}
\end{equation}
where $\Sigma$ is the mass density projected on the lens
plane. Note that this term depends only upon
the impact parameter $\vec\xi$, reflecting the fact that in
the thin lens approximation the effect of the gravitational
potential of the lens is dominated by that
part of the trajectory closest to the deflector. Here
$\xi _0$ is an arbitrary length scale.
Following Fermat's principle, we extremize the time of travel
with respect to $\vec\theta$ and arrive at the lens equation:
\begin{equation}
\vec\theta -\vec\beta -{1\over 2L}\
{\partial \psi\over\partial\vec\theta }=0\  . \label{EL}
\end{equation}
Notice that ${\partial \psi /\partial\vec\theta }= L
{\partial \psi /\partial\vec\xi }$.
The solutions to this equation give the angular positions of the
images.

Now we show the equivalence between the effect of a gravitational
wave and an effective lens
misalignment. Consider a lens geometry with
deflector, source, and observer aligned at $z=0,-L,L$ respectively, as
in Fig. 1, along the $z$-axis. Let U be the gravitational potential of
the deflector. Consider a gravitational wave propagating at an angle
$\vartheta$ with respect to the lens axis. We take the $(x,z)$ plane
as that containing the lens axis and the direction of propagation of
the gravitational wave. The metric perturbation caused by the wave
can be expressed as:
\begin{equation}
h_{ij}=\left(\begin{array}{ccc}
-\cos ^2\vartheta \ h_+ \ &-\cos\vartheta \ h_\times \ & \sin\vartheta
\cos\vartheta
\ h_+ \\
-\cos\vartheta \ h_\times &h_+&\sin\vartheta \ h_\times \\
\sin\vartheta \cos\vartheta \ h_+&\sin\vartheta \ h_\times &-\sin
^2\vartheta
\ h_+
\end{array}
\right)
\cos (\omega t-\vec k\cdot\vec x)\label{H}
\end{equation}
with propagation vector $\vec k=\omega (\sin\vartheta ,0,\cos\vartheta )$,
and $h_+$ and $h_\times $ the amplitudes of the two wave
polarizations. The total metric is then given by:
\begin{equation}
ds^2=(1+2U)dt^2-(1-2U)(dx^2+dy^2+dz^2)+h_{ij}dx^i dx^j \  . \label{DS}
\end{equation}
Along a null path, the time of travel is given by
\begin{equation}
T\approx \int _{-L}^L \ dz \left[1+{1\over 2}\left({dx\over dz}\right)^2
+{1\over 2}\left({dy\over dz}\right)^2
+{1\over 2}h_{ij}{dx^i\over dz} {dx^j\over dz}-2U \right]\  . \label{TTT}
\end{equation}
Notice that $\int 2U\ dz$ is the same as what we had previously
defined by $\psi$ in eq.\ (\ref{T1}).

We now define a family of null trial paths along which we will
integrate eq.\ (\ref{TTT}). Each path is built out of two
segments deflected by an angle $\vec\alpha$ at the deflector
plane. Instead of taking straight trajectories, we let each segment
be a solution of the geodesic equations in the presence of the
gravitational wave, neglecting the potential $U$
of the deflector, since its effects are later taken into account
through Fermat's principle. The condition that a photon from the
source at $(x,y=0,z=-L)$ reaches the observer at $(x,y=0, z=L)$
defines a one-parameter family of
trajectories parametrized by an arbitrary vector
$\vec\epsilon=(\epsilon_x,\epsilon_y)$:
\begin{eqnarray}
{dx\over dz}&&=\epsilon_x
-{h_+\over 2}\sin\vartheta (1-\cos\vartheta )
\cos\omega (t_e+L+z(1-\cos\vartheta ))\nonumber\\
{dy\over dz}&&=\epsilon_y
+{h_\times }\sin\vartheta \cos\omega (t_e+L+z(1-\cos\vartheta ))
\quad\quad{\rm if}\quad z<0\quad , \nonumber\\
{dx\over dz}&&=-\epsilon_x+{h_+\over 2\omega L} \sin\vartheta
[\sin\omega (t_e+L(2-\cos\vartheta ))-\sin\omega
(t_e+L\cos\vartheta )]
\nonumber\\
&&\quad\quad-{h_+\over 2}\sin\vartheta (1-\cos\vartheta )
\cos\omega (t_e+L+z(1-\cos\vartheta ))\nonumber\\
{dy\over dz}&&=-\epsilon_y-{h_\times \over \omega L}
{\sin\vartheta \over (1-\cos\vartheta)}
[\sin\omega (t_e+L(2-\cos\vartheta ))-\sin\omega
(t_e+L\cos\vartheta )]
\nonumber\\
&&\quad\quad+{h_\times }\sin\vartheta  \cos\omega (t_e+L+z(1-
\cos\vartheta ))
\quad\quad{\rm
if}\quad z>0\  .\label{dxdy}
\end{eqnarray}
The wave also affects the apparent position of the deflector; as
before, we  change variables from $\vec\epsilon$ to
the relative angular position between the
image and the deflector at the observer's position,
which we denote by $\vec\theta$. We find the relation (assuming
$\omega L\theta <<1$):
\begin{equation}
\vec\theta = (\epsilon_x, \epsilon_y)+
\left({h_+\ \sin\vartheta \over 2\omega L},{-h_\times \ \sin\vartheta
\over \omega L(1-\cos\vartheta )}\right)
[\sin\omega (t_e+L \cos\vartheta )-\sin\omega (t_e+L)]
\  . \label{THETA}
\end{equation}
The deflection $\vec\alpha$ imprinted upon the trajectory at the
lens plane $z=0$ can be written as
\begin{equation}
\vec\alpha =2(\vec\theta -\vec\beta_g) \label{ALPHA}
\end{equation}
if we define $\vec\beta_g$  as
\begin{eqnarray}
\vec\beta_g\equiv&&
\left({h_+\over 4\omega L}{\sin\vartheta },{-h_\times\over 2\omega L}
{\sin\vartheta
\over (1- \cos\vartheta )}\right)\nonumber\\
&&\times [\sin\omega (t_e+L(2-\cos\vartheta ))+
\sin\omega (t_e+L \cos\vartheta )-2\sin\omega
(t_e+L)]\nonumber\\
=&& -\left({h_+\over \omega L}{\sin\vartheta },{-h_\times\over\omega L}
{2\sin\vartheta
\over (1- \cos\vartheta )}\right)
\sin ^2 \left[{\omega L\over 2}(1-\cos\vartheta )\right]\sin\omega (t_e+L)
\label{BETA}
\end{eqnarray}
Of course, eqs. (\ref{THETA}) and (\ref{BETA}) reduce to our
previous eqs. (\ref{theta}) and
(\ref{beta}) in the case $\vartheta =\pi /2, h_\times =0$.

Now we are ready to find the time of travel, integrating eq.\
(\ref{TTT}).
One important thing to
note is that, for a thin lens, integration
of the last term in\ (\ref{TTT}), the contribution of the deflector
gravitational potential $U$, gives $-\psi(\vec\xi)$, where
$\vec\xi=\vec x(z=0)$ is the impact parameter of the trajectory.
And, if $\omega L\theta <<1$, the relation
$\vec\xi=L\vec\theta$
still holds, as in the absence of a gravitational wave.
At the end, we find for the total time of travel:
\begin{equation}
T=2L +|\vec\theta|^2\ L-2\ \vec\beta_g\cdot \vec\theta \ L-
\psi (L\vec\theta)\  ,\label{TTTT}
\end{equation}
Since this has exactly the same functional dependence on
$\vec\theta$ as in eq.
(\ref{T1}), we confirm the equivalence between an aligned lens
in the presence of a gravitational wave and a lens with an effective
lack of alignment given in terms of the wave parameters by
$\vec\beta_g$ of eq.\ (\ref{BETA}).

In the special case of an axially symmetric lens and in the limit
$\vert\vec\beta\vert <<|\vec\theta|$, the solutions
$\vec\theta_{1,2}$ to the lens
equation lie in the plane that contains the lens axis and
the direction of $\vec\beta_g$. That plane forms an angle $\phi$
with the plane that contains the
gravitational wave propagation vector (which we took as the
$(x,z)$ plane) such that
\begin{equation}
\tan\phi={(\beta_g)_y\over (\beta_g)_x}=
-{2h_\times\over h_+(1-\cos\vartheta)}\ .\label{phi}
\end{equation}
The solutions are then of the form
\begin{equation}
\vec\theta_{1,2}\approx \pm\vec\eta + a\vec\beta\  ,\label{12ax}
\end{equation}
with
\begin{equation}
\vec\eta=\eta(\cos\phi,\sin\phi)\quad  ;\quad  \eta={1\over 2L}
{\partial \psi\over\partial\theta
}\bigg |_{\theta=\eta}\  \label{ELL}
\end{equation}
a solution to the unperturbed lens equation, and
$a$  a coefficient that depends upon the lens model:
\begin{equation}
a^{-1}=1-{1\over2L}{\partial^2\psi\over\partial\theta}
\bigg |_{\theta=\eta}\  .\label{A}
\end{equation}
For instance, $a=1/2$ for a Schwarzschild lens, where
$\psi=4GM\ln\theta$, and $a=1$ for a singular isothermal sphere,
where $\psi=4\pi\Sigma_v^2\theta$, with $\Sigma_v$ the
velocity dispersion. For a non-singular lens, there must be
an odd number of images; in that case, eq. (\ref{12ax}) refers, say,
to the outer two images (the central third image is usually
de-magnified).

The time of travel for these solutions can be expanded as
\begin{equation}
T_{1,2}=2L+(|\vec\eta|^2\pm 2a \vec\beta \cdot \vec\eta)L\mp
2\vec\beta \cdot \vec\eta L
-\psi (\pm L\vec\eta)-a\vec\beta\cdot{\partial
\psi\over\partial\vec\theta    }\bigg
|_{\vec\theta=\pm\vec\eta}\  . \label{TAP}
\end{equation}
Using the unperturbed lens equation\ (\ref{ELL}) we see that the
two contributions to the time
of travel proportional to $a\vec\beta$, one
of geometric origin and the other due to the deflector's gravitational
potential, cancel each other. Besides, $\psi(L\vec\eta)=\psi(-
L\vec\eta)$ for an axially symmetric lens. Finally, the  time delay
between two images for a thin, axially symmetric lens, in the limit
$|\vec\beta|<<|\vec\eta|$ and with $\omega L\eta<<1$ is
\begin{equation}
\Delta T\approx -4\vec\beta\cdot \vec\eta L
={4\eta\over\omega }\sin\vartheta \left[h_+
\cos\phi +h_\times {2\sin\phi\over (1-\cos\vartheta )}\right] \sin
^2 \left[{\omega L\over
2}(1-
\cos\vartheta )\right]\sin\omega (t_e+L)\ ,\label{DDT}
\end{equation}
where $\phi$ is the angle between the plane containing the
photon trajectories and the plane that contains the gravitational wave
propagation vector, as given by\ (\ref{phi}). In the limit
$\beta_g<<\eta$, expression\ (\ref{DDT}) agrees with that of
Ref. \cite{Allen89}, evaluated
through integration of the Sachs-Wolfe formula along unperturbed
photon paths.

\section{CONCLUSIONS}

 We have shown that the lens equation for a thin, axially aligned
gravitational lens configuration
in the presence of a very long wavelength
gravitational wave is equivalent to that of a similar lens with the
source out of alignment and no gravitational wave. An
observer who measures time delays, angular positions, or any other
observables such as relative magnifications and redshifts, and uses them
to reconstruct the lens configuration, cannot tell the two situations
apart. Thus, an observer ignorant of gravitational waves would naturally
and `correctly' interpret the observations as a simple non-aligned lens.
This conclusion is valid if $\omega L\eta <<1$, which is
the interesting range since the induced time
delays are largest when $\omega L\approx 1$. We performed our
calculations around an aligned lens configuration and with
source and observer equidistant from the deflector, but it is clear
that the conclusion holds in more general cases: the effect of a long
wavelength gravitational wave upon a lens with a given geometry is
equivalent, from the observer's viewpoint, to an effective change of
lens geometry. Consequently, measuring time delays in gravitational
lenses does not provide a  method for probing a cosmological background of
gravitational waves.

Formally, Allen's result for the wave-induced time delay is correct
in the small amplitude limit: for $\beta_g<<\eta$,
eq.\ (\ref{DDT}) for the time delay induced by a
gravitational wave upon a thin, axially symmetric lens
agrees with that of Ref. \cite{Allen89}.
In the opposite limit, $\beta_g \gg \eta$, the effect of the gravitational
wave is equivalent to a change in the alignment between source,
deflector, and observer by an amount that exceeds the typical
deflection angle the deflector can imprint, precluding the
formation of multiple images in the case of an aligned lens.
In this case, multiple images can only form if there is a compensating
geometric misalignment between source and deflector, and the geometric
delay will partially cancel the lens-induced delay.
Thus, even if
$h>>\eta$, the total time delay does not exceed the typical
intrinsic lens time delay of order  $L\eta^2$. Moreover, in either
limit, the measured time delay is just what the observer would
expect in the complete absence of gravitational waves, based on her
measurements of the lens observables.

The detection by the COBE satellite \cite{COBE} of a quadrupole
anisotropy in the cosmic microwave background  places a
bound  $h\alt 6\times 10^{-6}$ on the amplitude of cosmological
gravitational waves. If a large fraction of the anisotropy detected by
COBE is due to gravitational waves, a possibility that can be
accommodated by many inflationary cosmological models
\cite{Abbott84,COBEGW}, then the wave-induced
time delays between multiple images of quasars are
comparable to typical intrinsic lens time delays, with
$\eta\approx 10^{-5}$. One could have hoped that careful lens
modelling could allow one, at least in principle if not in practice,
to separate the wave-induced from the intrinsic time delay, and thus
reveal the presence of the gravitational waves.
Our work indicates that this is not a possibility.

\section{ACKNOWLEDGEMENTS}

The work of J.F. was supported by DOE and by NASA (grant NAGW-2381)
at Fermilab. He thanks B. Allen and M. Sasaki for helpful
conversations and especially thanks N. Kaiser and M. Turner for their
collaborations on aspects of this problem. The work of
D.H. and G.S. was supported by CONICET and by Fundaci\'on Antorchas.

\newpage
\begin{center}
\setlength{\unitlength}{0.67cm}
\begin{picture}(24,9)(-11.5,-4.5)
 \put(0,0){\circle*{0.8}}
\put(-10,0){\circle*{0.3}}
\put(10,0){\circle*{0.3}}
 \put(-11.5,0){\vector(1,0){24}}
\put(0,-4.5){\vector(0,1){9}}
 \thicklines
\put(-10,0){\line(4,1){10}}
\put(-10,0){\line(4,-1){10}}
\put(10,0){\line(-4,1){10}}
\put(10,0){\line(-4,-1){10}}
 \put(-11,0.5){\large source}
\put(9.5,0.5){\large observer}
\put(-10.2,-0.8){-L}
\put(9.7,-0.8){\ L}
\put(0.5,0.7){\large M}
\put(0.3,4){\it x}
\put(12,-0.7){\it z}
\put(4.9,0.4){\large $2 \eta $}
 \bezier{60}(6,-1)(5.5,0)(6,0.9)
  \end{picture}
\end{center}

\leftskip0.5cm\noindent
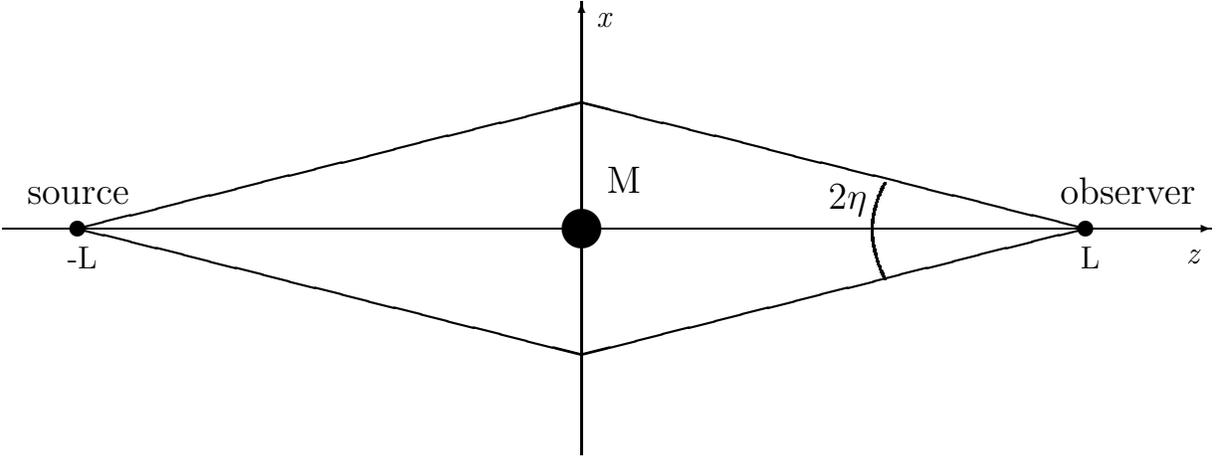
\underbar{Figure 1}: The geometry of an aligned lens. The deflector
is a point mass M at the origin of
coordinates.  Source and observer lie
along the {\it z}-axis, equidistant from the deflector.
The observer sees two images of the
same source (actually an Einstein ring)
with angular separation 2$\eta $. The trajectories are
approximated by straight segments.


\newpage

\begin{center}
\setlength{\unitlength}{0.67cm}
\begin{picture}(24,9)(-11.5,-4)
 \put(0,0){\circle*{0.8}}
\put(-10,1.666){\circle*{0.3}}
\put(10,0){\circle*{0.3}}
 \put(-11.5,0){\vector(1,0){24}}
\put(0,-4){\vector(0,1){9}}
 \thicklines
\put(-10,1.666){\line(6,1){10}}
\put(-10,1.666){\line(3,-1){10}}
\put(10,0){\line(-3,1){10}}
\put(10,0){\line(-6,-1){10}}
\bezier{30}(0,3.333)(-2.5,4.2)(-5,5)
 \bezier{110}(-10,1.666)(0,0.833)(10,0)
  \put(-11,2.2){\large source}
\put(9.5,0.5){\large observer}
\put(-10.2,-0.8){-L}
\put(9.7,-0.8){\ L}
\put(0.6,0.2){deflector}
\put(0.3,4.5){\it x}
\put(12,-0.7){\it z}
 \bezier{10}(-10,1.666)(-10,0.833)(-10,0)
  \put(-3.4,0.3){$\vec\beta$}
\put(4.3,0.9){$\vec\theta_2$}
\put(3,-0.7){$\vec\theta_1$}
\put(0.25,2){$\vec\xi $}
\put(-2.5,3.33){$\vec\alpha $}

\bezier{40}(-2.6,0)(-2.6,0.5)(-2.5,1.03)
\bezier{40}(5.1,0)(5.1,1)(5.28,1.5)
\bezier{40}(3.8,0)(3.8,-0.5)(3.9,-1)
\bezier{40}(-1.85,3.05)(-1.95 ,3.43)(-1.75,3.9)
 \end{picture}
\end{center}

\leftskip0.5cm\noindent
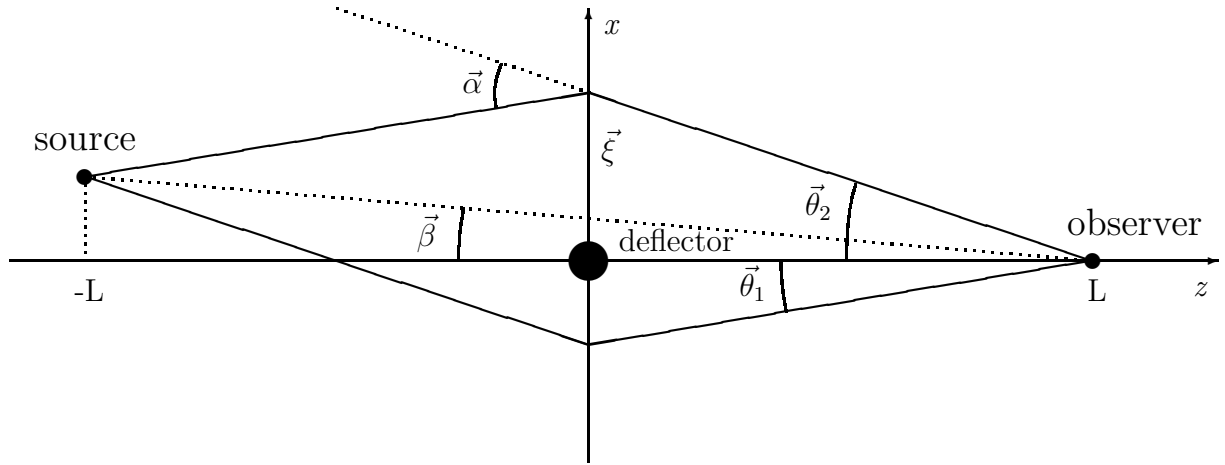
\underbar{Figure 2}: The geometry of a non-aligned lens. The source
forms an angle
$\vec\beta$ with respect to the line that joins observer and deflector.
$\vec\theta_1$ and $\vec\theta_2$ are the angular positions of the
images, $\vec\xi $ is the impact
parameter, and $\vec\alpha $ the deflection angle.

\end{document}